\documentclass[letter]{aa} 

\usepackage{txfonts}
\usepackage[colorlinks,citecolor=blue]{hyperref}
\usepackage{graphicx}   
\graphicspath{{images/}}

\begin{document}


\title{First direct identification of the barlens vertical structure in galaxy models}

\author{
  Iliya S. Tikhonenko\inst{1}
\and 
  Anton A. Smirnov\inst{1,2}
\and 
  Natalia Ya. Sotnikova\inst{1}
}
\institute{
  St. Petersburg State University,
  Universitetskij pr.~28, 198504 St. Petersburg, Stary Peterhof, Russia \\
  \email{n.sotnikova@spbu.ru}
  \and
  Central (Pulkovo) Astronomical Observatory of RAS, Pulkovskoye Chaussee 65/1, 196140 St. Petersburg, Russia
}


\date{Received XXX; accepted YYY}


 






\abstract
{
Applying spectral dynamics methods to one typical $N$-body model with a barlens, we dissect the modelled bar into separate components supported by completely different types of orbits. We identify at least four components: a narrow elongated bar, a boxy bar, and two components contributing to the barlens. We analyse the vertical structure of all components that make up the thick part of the bar, which has a boxy/peanut shape (B/P bulge). We show that the `peanut' shape is mainly due to the orbits that assemble the boxy part of the face-on bar. We associate the X-shape with the narrow and elongated bar. The wider part of the barlens with square-like isophotes contributes to the boxy shape of the B/P bulge when we observe the galaxy edge-on. However, the part of the barlens with rounded isophotes in the face-on view is a rather flat structure in the vertical direction without any significant off-centre protrusions. Thus, for the first time, we demonstrate that the rounded face-on barlens cannot be entirely associated with the B/P bulge.
}


\maketitle

\section{Introduction} 
\par
About half of the edge-on galaxies possess boxy/peanut-shaped (B/P) bulges (\citealp{Lutticke_etal2000}, but see \citealp{Yoshino_Yamauchi2015}). In some cases, these bulges can exhibit an X-shape, therefore, the abbreviation B/P/X is sometimes used for such structures. 
The fractions of B/P bulges and bars in galaxies of Hubble types earlier than Scd are almost the same (e.g. \citealp{Erwin_Debattista2017} and \citealp{Li_etal2017}).
It is widely believed that B/P bulges are the inner parts of bars that have thickened in the vertical direction (e.g. \citealp{Laurikainen_Salo2016} and  \citealp{Athanassoula2016}, and references therein). 
Such a connection follows, first of all, from the results of $N$-body simulations which show that stellar discs are often subject to the bar-like instability. Subsequently, the evolution of the bar is accompanied by the formation of a vertically thin outer structure and a vertically thick inner structure (\citealp{Combes_Sanders1981} and \citealp{Raha_etal1991}). Viewed side-on (along the minor axis of a bar), the thick inner part of a bar appears as a peanut-shaped stellar structure. At intermediate position angles of a bar, this structure shows a boxy-like shape.
\par
Although B/P bulges are usually detected in the edge-on view, in some works it is accepted that they are counterparts of the so-called barlenses \citep{Laurikainen_etal2011}, which are lens-like structures embedded in narrow and elongated bars and seen rather face-on (see the review by \citealp{Laurikainen_Salo2016}). An understanding of the common nature of B/P bulges and barlens components came from an analysis of $N$-body simulations. \citet{Athanassoula_etal2015} viewed a set of $N$-body snapshots with a barlens from different viewing angles and conclude that a barlens is the vertically thick part of the bar but viewed face-on, that is to say a barlens and a B/P bulge are the same part of the bar, but simply viewed from a different viewing angle.
\par
The main proof that the barlens is another way of thinking about the projected B/P bulge is the coincidence of their relative sizes. From the photometric analysis of nearly face-on galaxies, \citet{Erwin_Debattista2013,Erwin_Debattista2017} {find} that B/P features extend from one-third to two-thirds of the bar radius, while \citet{Laurikainen_etal2011}, \citet{Athanassoula_etal2015}, and \citet{Herrera-Endoqui_etal2015} {obtain} similar results using barlenses in face-on galaxies. Using a stellar-kinematic diagnostic, \citet{Mendez-Abreu_etal2008} {measure} the location of vertical protrusions in the barlens galaxy {NGC~98} at about one-third of the bar's radius. 
All these pieces of evidence led to the view that a barlens and a B/P bulge can be {regarded as} the same structure, but viewed from different angles.
\par
\citet{Erwin_Debattista2017} express a slight doubt about such a coherent picture. First, it is not clear if barlens-like morphologies are always due to a B/P structure. In addition, the definition of a `lens' is somewhat ambiguous. It is unclear whether lenses and barlenses can be considered as similar structures. Some authors define the lens as a shallow feature on the brightness profile with a further steeper declination \citep{Kormendy_1979}, while others identify the barlens by the steep quasi-exponential profile \citep{Laurikainen_etal2014}. Some additional argumentation concerning barlens identification can be found in Appendix~A of \citet{Erwin_Debattista2017}. 
\par
Another major problem is that the thick part of the bar is a complex structure. Based on the spectral analysis of individual orbits of stars from different $N$-body models, \citet{Smirnov_etal2021} dissected the modelled bars into separate components that are supported by completely different orbital groups. They identified at least four orbital components: the {narrow elongated bar}, the face-on boxy bar, and two components constituting the barlens. All these components may have their unique vertical structure and a prominent and large face-on component does not have to be thick in the edge-on view. Moreover, \citet{Smirnov_etal2021} determined that these components differ rather little in their linear sizes, but they have a rather different morphology. Thus, the coincidence of relative sizes also does not serve as an argument in favour of the structure coincidence. Therefore, observing a bar at different viewing angles does not help to differentiate all its components. Dynamical arguments are required to reliably separate the components from each other. \citet{Smirnov_etal2021} focus on the face-on morphology of bars. In this letter, we consider an $N$-body model with a barlens (Sect.~\ref{sec:model}) and analyse the edge-on structure of all the components that make up the bar (Sect.~\ref{sec:structure}). 
In Sect.~\ref{sec:conclusions} we come to conclusions about the vertical structure of all identified components of the bar.

\section{$N$-body model and orbital frequencies}
\label{sec:model}

We analyse an $N$-body model of a barred galaxy that evolved from an initially axisymmetric and equilibrium state with no gas involved. The model initially consists of an exponential disc isothermal in the vertical direction embedded in a live dark matter halo. The halo was modelled by a truncated sphere with the density profile close to the Navarro-Frenk-White (NFW) profile \citep{NFW}. During its evolution, the disc naturally forms a bar. To obtain the barlens in simulations, one can introduce an additional component that increases the central mass concentration in the initial model. This connection between the barlens and the central mass component was previously established in the following studies. In  simulations by \citet{Athanassoula_etal2013}, it is the increase in the gas concentration in the central area via the bar-induced inflow. At the same time, \citet{Salo_Laurikainen2017} conclude that the introduction of a  small and compact classical bulge in the N-body model drastically changes the bar morphology and leads to the formation of a central barlens in a face-on view. In the present paper, we analyse one of the models by \citet{Smirnov_etal2021} with a small compact bulge (labelled as a BL model in their paper), which demonstrates a barlens in its structure.
\par
The details of numerical modelling can be summarised as follows.
We consider a dynamically cool disc with a Toomre parameter value of $Q=1.2$ at $R=2R_\mathrm{d}$, where $R_\mathrm{d}$ is the disc radial scale length. We scaled the model to obtain a Milky Way-like galaxy and assumed $R_\mathrm{d}=3.5$ kpc and $M_\mathrm{d}=5 \cdot 10^{10} M_\mathrm{\sun}$, where $M_\mathrm{d}$ is the disc mass. Then the time unit will be $t_\mathrm{u}\approx 14$ Myr.
The parameters of the dark halo were chosen to produce a reasonable dark halo profile with $M_\mathrm{h}(r<4R_\mathrm{d})/M_\mathrm{d} \approx 1.5$. The model possesses a classical bulge with a \cite{Hernquist1990} density profile. The total mass of the bulge is $M_\mathrm{b}=0.1 M_\mathrm{d}$,  with the scale parameter $r_\mathrm{b}=0.05 R_\mathrm{d}$.
\par
We used 4$kk$ particles for the disc, 4.5$kk$ for the halo, and 0.4$kk$ for the bulge. The evolution of the model was followed up to 8 Gyr. 
\par
\begin{figure}
  \centering
  \includegraphics{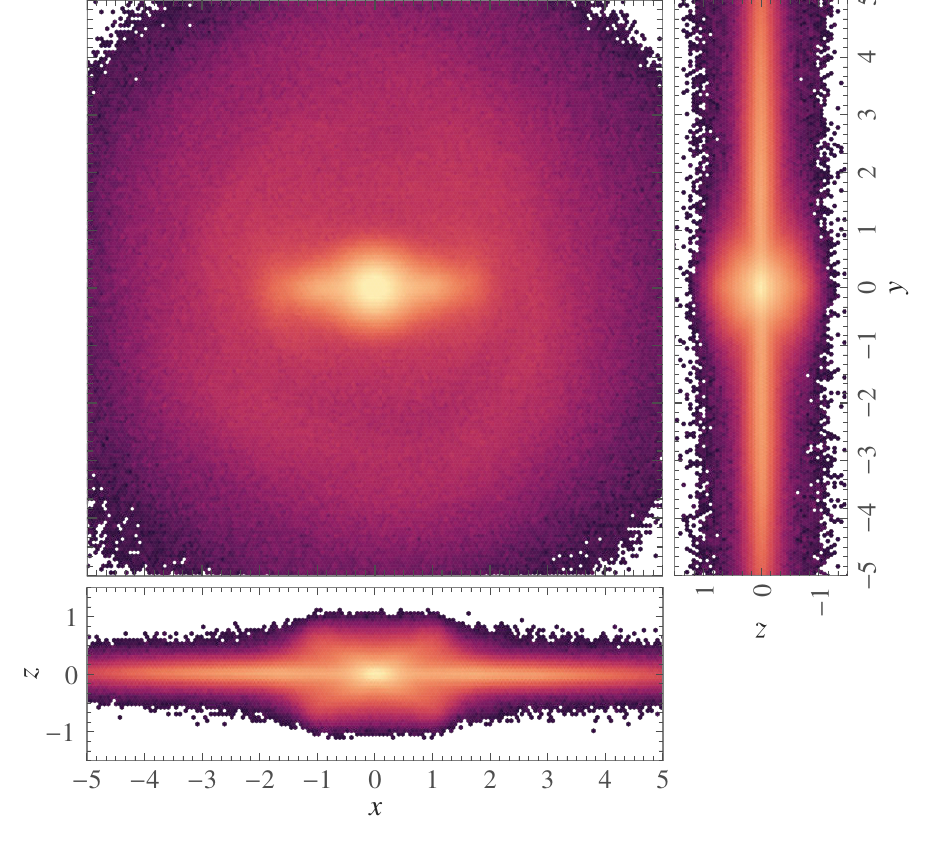}
\caption{Face-on, side-on, and end-on views of the considered model at $t=450$ (6 Gyr).}
  \label{fig:modeloverview}
\end{figure}

To study the orbital composition of the bar, we applied the methods of spectral dynamics~\citep{Binney_Spergel1982} and obtained orbital frequencies in the evolving {bar and disc}. As the pattern speed varies, we estimated the frequency shifts and found that for the time interval $t=400\text{--}500$ ($5.5$--$7$ Gyr, when the bar pattern speed is more or less established), the shift is small for most of the particles, and its value is about the frequency measurement error \citep{Parul_etal2020}. We chose this time interval to apply the frequency analysis to all disc particles and obtained $f_x$, $f_y$, $f_z$, and $f_\mathrm{R}$, which are frequencies of oscillations along $x$, $y$, and $z$ coordinates and cylindrical radius $\mathrm{R}$. In Fig.~\ref{fig:modeloverview} $xy$, $xz$, and $yz$ snapshots for the whole BL model are presented at the time moment $t=450$. A barlens is clearly seen in the $xy$ snapshot, while a thick B/P structure is visible in the $xz$ view. 

\section{Edge-on structure}
\label{sec:structure}

\begin{figure*}[h]
  \centering
  \includegraphics[width=\linewidth]{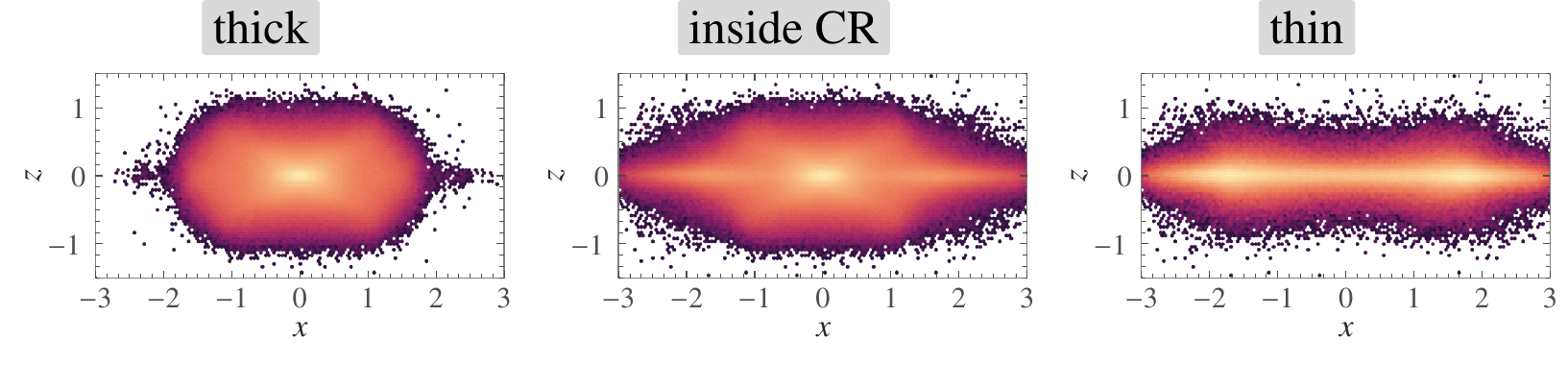}
  \caption{Side-on views of particles in the considered model at $t=450$ (6 Gyr).
  \emph{Left:} Vertically thick part of the bar. \emph{Middle:} All particles inside the co-rotation. \emph{Right:} Remaining particles which form a vertically thin structure.}
  \label{fig:BPSdisksep}
\end{figure*}
\par
\begin{figure}
  \centering
  \includegraphics[width=0.9\linewidth]{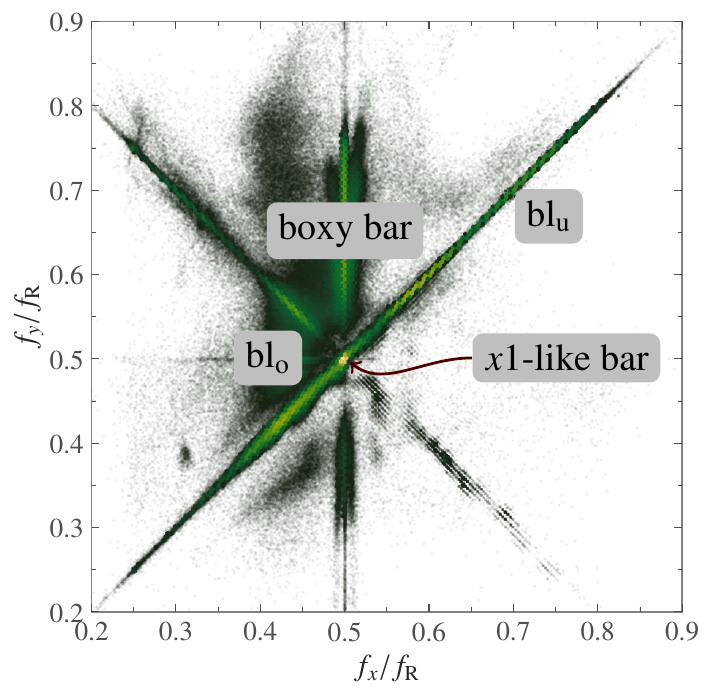}
  \caption{Distribution of all orbits in the bar over the ratios of dominant frequencies $f_x/f_R$ and $f_y/f_R$ for the considered model at $t=450$ (6 Gyr). The annotations illustrate the separation of orbits into orbital groups.}%
  \label{fig:familysep}
\end{figure}

In \citet{Smirnov_etal2021}, the BL model was disassembled into several orbital groups to analyse the face-on morphology of an overall bar. We assume that each individual group is characterised by a unique set of frequency ratios $f_\mathrm{R}/f_x$ and $f_\mathrm{R}/f_y$. In \citet{Smirnov_etal2021}, we focus on the face-on morphology of the inner parts of bars in our models. For this purpose, at the initial stage of our orbital analysis, we discarded orbits from the rest of the disc (outside the ring at the co-rotation radius (CR) $R_\mathrm{c} \approx 3$) as well as of the thin part of the bar itself. Orbits from the thin part of a bar are located inside the CR and can be distinguished by the frequency ratio $f_z/f_x>2.25$ \citep{Smirnov_etal2021}. Although the fraction of the 4$kk$ particles in these orbits is about 14\%, all these orbits do not contribute to the thick B/P bulge, but only to the most remote and vertically thin parts of the bar (see Fig.~\ref{fig:BPSdisksep}). 
For the thick part of the bar, we identified all separate orbital groups (see Fig.~\ref{fig:familysep}) based on frequency ratios $f_x/f_\mathrm{R}$ and $f_y/f_\mathrm{R}$ and show that the structures to which they contribute are completely different in morphology. In this letter, we analyse the edge-on morphology of the thick bar and its components in our BL model.
\par
\citet{Smirnov_etal2021} identified the following components
constituting a complex bar with a barlens: a `classic' elongated bar ($f_\mathrm{R}/f_x=2.0\pm0.1$, 21.8\% of particles from 4$kk$) and two components contributing to the barlens (\hbox{$\text{bl}_{\text{u}}$}\xspace orbits, in our notation, with $f_\mathrm{R}/f_x\leq 1.9$, 9.2\%, and \hbox{$\text{bl}_{\text{o}}$}\xspace orbits, with $f_\mathrm{R}/f_x\geq 2.1$, 13.5\%). A `classic' bar {can be further divided} into two subsystems. One is assembled from the x1-like orbits
($f_y/f_x=1.0 \pm 0.1$, 11.4\%), and another one is the so-called boxy bar with orbits that are characterised by the ratio $f_y/f_x\geq 1.1$ (10.4\%). In Fig.~\ref{fig:familysep}, all these orbital groups occupy separate regions (a spot and straight rays) and can be easily distinguished.
\par
\begin{figure}
  \centering
  \includegraphics[width=\linewidth]{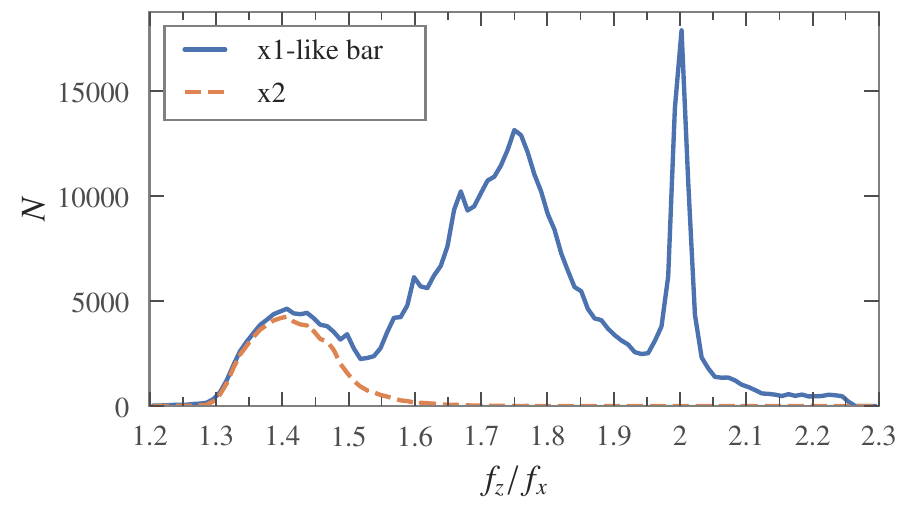}
  \caption{Distribution of all orbits in the x1-like bar over the $f_z/f_x$ ratio for the considered model at $t=450$ (6 Gyr). The same distribution
  is indicated with the different colour for the inner bar supporting family, which makes up most of the peak at $1.4$ in the primary distribution.}%
  \label{fig:x1likebarsubsets}
\end{figure}

The x1-like bar has a complex structure itself. Three main orbital blocks, which contribute to morphologically different structures, can be distinguished. Few orbits are parented from the so-called x2 orbital family (1.7\%). At the distribution over the ratio $f_z/f_x$ (Fig.~\ref{fig:x1likebarsubsets}), these orbits have a wide peak at $f_z/f_x=1.4$.  
Two other subsystems have a hump at $f_z/f_x=1.75$ (orbits elongated along the bar) and a very narrow peak at $f_z/f_x=2.0$. The last group consists of `banana'-like orbits.
\par
\begin{figure*}
  \centering
  \includegraphics{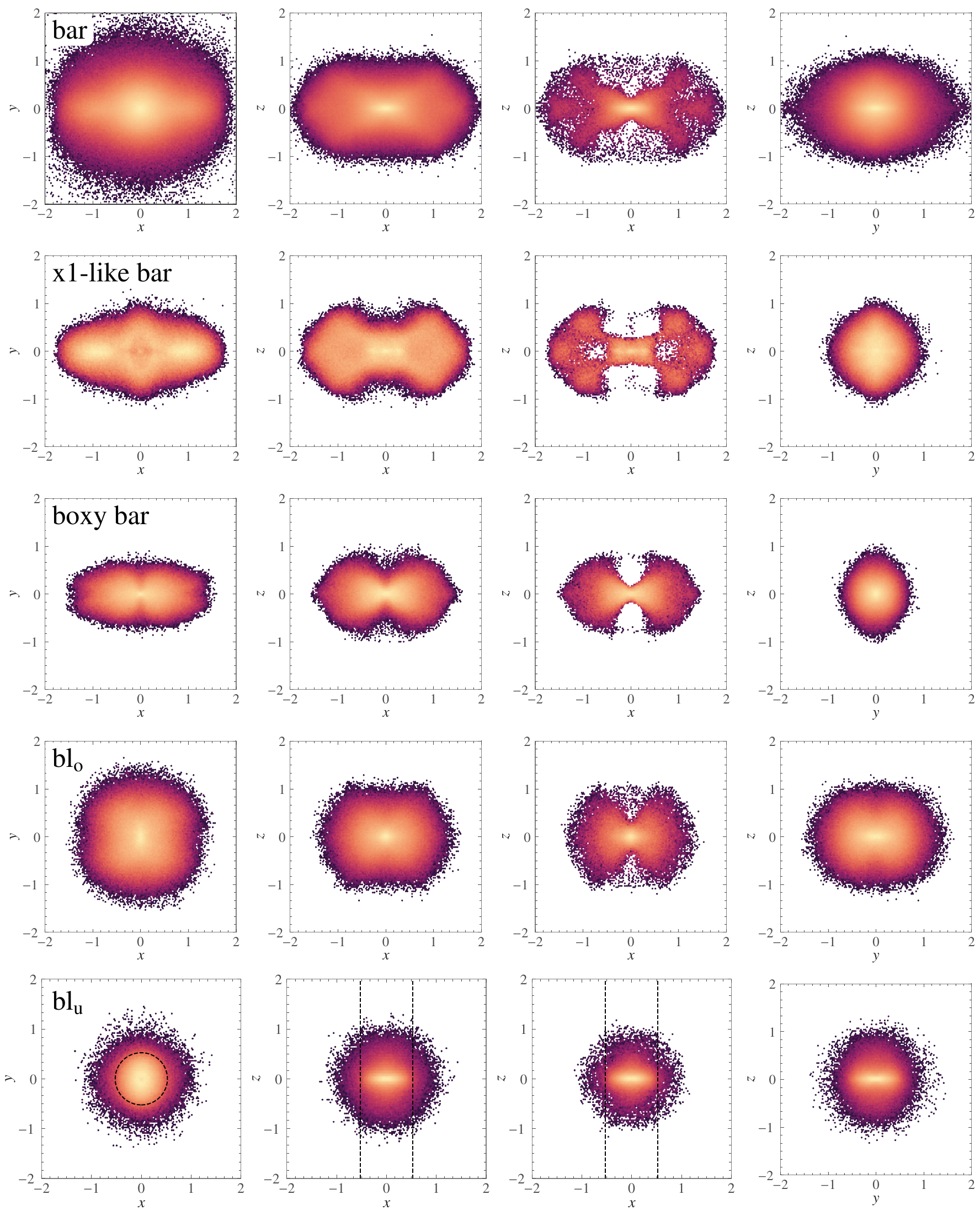}
  \caption{Bar building blocks in different projections. From left to right: Projection in the $xy$ plane, in the $xz$ plane, unsharp-masked image of 
  $xz$ projections, and $yz$ projections. The annotations on the $xy$ views in the first column indicate the name of the corresponding orbital group for the whole row. Dashed black lines in the last row show the value of two half-mass
  radii for the $\hbox{$\text{bl}_{\text{u}}$}\xspace$ component.
}%
  \label{fig:xz_collage}
\end{figure*}

Fig.~\ref{fig:xz_collage} (the upper row) demonstrates face-on ($xy$), side-on ($xz$), and end-on ($yz$) views of a modelled B/P bulge (Fig.~\ref{fig:xz_collage}; the upper row) and its orbital subsystems (x1-like bar, boxy bar, and \hbox{$\text{bl}_{\text{o}}$}\xspace and \hbox{$\text{bl}_{\text{u}}$}\xspace components) in our BL model. We also provide unsharp-masked plots\footnote{We created unsharp-masked plots following the scheme outlined by \citet{Bureau_etal2006}.} of $xz$ snapshots (third column) to highlight the vertical structural features of the bar and its components. 
\par
It can be seen that the peanut-like structure and the X-structure in the $xz$ view (Fig.~\ref{fig:xz_collage}; second column) are created by boxy and x1-like parts of the `classic' bar. The peanut-like structure is more inherent in the boxy part of the bar, while the x1-like bar demonstrates four bright rays of the X-structure\footnote{At the initial stages of evolution, the boxy bar also exhibits four noticeable rays, but over time they wash out and a peanut-like structure remains (see the third column in Fig.~\ref{fig:xz_collage} with unsharp-masked plots).}, which outline the $z$-maxima of various quasi-periodic orbits \citep{Parul_etal2020}.
\par
Three orbital blocks of the x1-like bar demonstrate different structural features seen at $xy$ and $xz$ projections. First of all, the $xy$ projection clearly exhibits the existence of an inner perpendicular bar (the orbits around the x2 family; they have a wide peak at $f_z/f_x=1.4$). In the unsharp-masked $xz$ plot for the `x1-like' orbital group, this structure reveals itself as a flat subsystem. 
Banana-like orbits ($f_z/f_x=2.0\pm0.05$) delineate the most remote parts of a bar (e.g. \citealp{Valluri_etal2016} and \citealp{Patsis_Athanassoula2019}). They appear as two distant shell-like features in the $xz$ unsharp-masked plot for the x1-like component. As to the orbits elongated along the bar and having a hump at $f_z/f_x=1.75$, they are the main contributors to the X-structure and bright rectilinear rays protruding from the main plane of the disc \citep{Parul_etal2020}.  
\par 
The orbits from the \hbox{$\text{bl}_{\text{o}}$}\xspace group (a part of a face-on lens) are assembled into a smooth square-like structure in all three projections with slight traces of a `peanut' in the central region (see an unsharp-masked plot in Fig.~\ref{fig:xz_collage}).
The whole structure resembles a 3D cube with rounded corners and dents in the middle of the cube faces. 
If there are many such orbits, then, at intermediate P.A. of the bar, the structure assembled from such orbits can be responsible for the box-like isophotes of the B/P bulge because this structure retains its cube-like shape in all projections, while other orbital subsystems develop a rounded shape. It is important to note that this subsystem is observed even in models without a classical bulge, but in such models the orbits inhabiting it are few (see, for example, \citealp{Gajda_etal2016}). 
\par
The most striking thing is that the \hbox{$\text{bl}_{\text{u}}$}\xspace orbits (the main part of a lens as is argued in \citealp{Smirnov_Sotnikova2018}) constitute the rounded structure not only in the face-on view (Fig.~\ref{fig:xz_collage}), but also in the $xz$ projection as well as in the $yz$ projection. The whole structure resembles a ball-shaped fruit with a flat bone inside (or a disc with a smooth halo). This is especially evident in the unsharp-masked plot. The flat part and the halo around it stand out. \citet{Smirnov_etal2021} show that the \hbox{$\text{bl}_{\text{u}}$}\xspace orbits are the main contributors to the rounded face-on barlens in models with compact classical bulges. The orbits identified have a rosette-like shape in the frame co-rotating with the bar. Such a shape is typical for orbits trapped around stable loop orbits from the nearly round families x2 or x4. The x2 orbits lie closer to the centre and rotate in the same direction as the bar in the inertial reference frame, and the x4 orbits rotate retrogradely, have a more rounded shape, and are located further from the centre. Examples of \hbox{$\text{bl}_{\text{u}}$}\xspace orbits can be found in \citet{Smirnov_etal2021} (their figure~14). They are not planar, but have a rather small vertical extension and create a flat structure. If such orbits dominate in the central regions, they should manifest themselves as nuclear discs with kinematics which are characterised by near-circular orbits \citep{Gadotti_etal2020}. In a subsequent study, we intend to go deeper into issues related to the kinematics of individual subsystems of B/P bulges. A smooth spherical halo from such orbits is also gradually formed, but no signs of the off-centre vertical protrusions are observed. In the last row of Fig.~~\ref{fig:xz_collage}, the dashed black circle shows the location of the two half-mass area for this substructure. In the $xz$ projection, this area is indicated by two vertical lines. On the graph one can then see that the thickest parts of the boxy and x1-like bars fall into the boundary of this area or lie outside it. Such a similarity of sizes can lead to a false identification of the barlens with the B/P bulge, although these are completely different structures.
\par

A barlens in the central regions of the galaxy can coexist with a B/P bulge, which manifests itself at a greater distance from the centre.  It comes as no surprise that nine of the galaxies where \citet{Erwin_Debattista2017} found B/P bulges also possess barlenses as reported by \citet{Laurikainen_etal2011}. For galaxies at intermediate inclinations, \citet{Erwin_Debattista2017} suggest the  photometric method for separating the thick part of the bar (B/P bulge) from the extended thin bar based on the isophote twisting. The analysis of isophotes makes it possible to identify the central barlens as well. If it is round and vertically thin, then the isophotes for intermediate inclinations will be elliptical and aligned with the elliptical isophotes of the outer disc. A perfect example is NGC~3992, where aligned ellipses are visible in the centre, and then there is an isophote twisting \citep[][figure~3]{Erwin_Debattista2017}. However, such an analysis for the models used is beyond the scope of the present letter and will be discussed elsewhere.

\par

\section{Conclusions}
\label{sec:conclusions}

We carry out the spectral analysis of one typical $N$-body model with a barlens. The analysis allows us to separate different types of orbits at the centre of galaxies. Our conclusions regarding the vertical structure of all identified groups are as follows.
\par
1. For the first time, we have revealed the vertical structure of the bar component that appears as a barlens in the face-on view.
In our $N$-body model the rounded barlens is a rather flat structure without signs of the vertical peanut shape. This part of the barlens has nothing to do with the B/P bulge. The coincidence of the barlens and B/P bulge sizes, which is often used to show the identity of these two structures, is rather accidental. As demonstrated by \citet{Smirnov_etal2021}, boxy orbits, which are assembled in a 3D `peanut', are shortened with the transition to models with a compact classical bulge. Both the barlens and the B/P bulge are shorter than the elongated bar, but these components are not different projections of the same structure.
\par
2. Apparently, the boxy shape of the B/P bulge is determined by \hbox{$\text{bl}_{\text{o}}$}\xspace orbits which are assembled in a 3D cube-like structure with slight traces of a `peanut'.
\par
3. The `peanut' component and the X-structure are mainly due to x1-like and boxy orbits. 
The `peanut' component (boxy bar) is the main contributor to the B/P structure and this component is present even in the models without classical bulges and barlenses. 
Thus, the B/P bulge is a barlens-independent component of the bar vertical structure, and its presence in the edge-on view is not an indicator of the obligatory presence of a face-on barlens.
\par
A detailed spectral analysis of the barlens model leads to the conclusion that B/P bulges and barlenses are complex structures consisting of morphologically different orbital subsystems. Therefore, more work is needed to understand their origins. In a forthcoming paper, we intend to examine the orbital structures using the kinematic data and offer tests to distinguish them.

\begin{acknowledgements}

The authors express gratitude for the grant of the Russian Foundation for Basic Researches number 19-02-00249. 
We thank the anonymous referee for his/her review and appreciate the comments, which contributed to improving the quality of the article.

\end{acknowledgements}

\bibliographystyle{aa}
\bibliography{article5}

\label{lastpage}

\end{document}